\documentclass[9pt,twocolumn,twoside]{osajnl}
\usepackage{txfonts}
\usepackage{physics}
\usepackage[detect-weight=true, detect-family=true]{siunitx}

\journal{ol} % Choose journal (ao, aop, josaa, josab, ol, pr)
\setboolean{shortarticle}{true}

\title{Resonant dispersive wave emission in hollow capillary fibers filled with pressure gradients}

\author[1, *]{Christian Brahms}
\author[1]{Federico Belli}
\author[1]{John C. Travers}

\affil[1]{School of Engineering and Physical Sciences, Heriot-Watt University, Edinburgh, EH14 4AS, UK}

\affil[*]{Corresponding author: c.brahms@hw.ac.uk}

\doi{\url{}}

\begin{abstract}
Resonant dispersive wave (RDW) emission in gas-filled hollow waveguides is a powerful technique for the generation of bright few-femtosecond laser pulses from the vacuum ultraviolet to the near infrared. Here we investigate deep-ultraviolet RDW emission in a hollow capillary fiber filled with a longitudinal gas pressure gradient. We obtain broadly similar emission to the constant-pressure case by applying a surprisingly simple scaling rule for the gas pressure and study the energy-dependent dispersive-wave spectrum in detail using simulations. We further find that in addition to enabling dispersion-free delivery to experimental targets, a decreasing gradient also reduces the pulse stretching within the waveguide itself, and that transform-limited pulses with \SI{3}{\fs} duration can be generated by using short waveguides. Our results illuminate the fundamental dynamics underlying this frequency conversion technique and will aid in fully exploiting it for applications in ultrafast science and beyond.
\end{abstract}

\setboolean{displaycopyright}{false}

\begin{document}

\maketitle
Laser sources delivering bright, wavelength-tuneable few-femtosecond pulses are a key goal in ultrafast photonics with wide-ranging applications, for instance in time-resolved spectroscopy \cite{kotsina_ultrafast_2019, maiuri_ultrafast_2020}. Resonant dispersive wave (RDW) emission in gas-filled hollow-core waveguides is a particularly promising approach to achieving this goal, offering wide, continuous tuneability from the vacuum ultraviolet \cite{ermolov_supercontinuum_2015, travers_high-energy_2019} to the near infrared \cite{brahms_high-energy_2019-1} with high conversion efficiency and few-femtosecond duration. This technique is based on soliton self-compression, which occurs when anomalous group-velocity dispersion continually compensates for the chirp induced by self-phase modulation during nonlinear propagation of a laser pulse. It was first demonstrated in small-core anti-resonant fibers at the microjoule pump energy scale \cite{joly_bright_2011}. Recently, we demonstrated that the use of large-core hollow capillary fibers (HCF) allows for significant energy scaling \cite{travers_high-energy_2019,brahms_high-energy_2019,brahms_high-energy_2019-1} while eliminating the tuneability gaps inherent to anti-resonant fibers \cite{mak_tunable_2013}. To make the potential offered by RDW-based sources a reality, a full understanding of the complex dynamics underlying the frequency conversion is required. Furthermore, several technical challenges, most importantly the delivery of compressed RDW-generated laser pulses to an experimental target, have to be overcome.

Here we study resonant dispersive wave emission in hollow capillary fibers under the influence of a longitudinal gas pressure gradient along the waveguide. Increasing pressure gradients, where the entrance of the HCF is under vacuum, are commonly used when compressing high-energy laser pulses by post-compression---spectral broadening in gas-filled HCF with subsequent phase compensation by dispersive optics---to improve the coupling to the waveguide \cite{nisoli_generation_1996,suda_generation_2005}. For RDW emission, decreasing gradients are of particular interest since they allow for dispersion-free delivery of generated pulses into a vacuum system \cite{brahms_direct_2019}. This is critical for experiments which exploit the few-femtosecond pulse duration of dispersive-wave pulses, especially in the deep and vacuum ultraviolet. In this spectral region, broadband dispersion compensation is unavailable, and the only way to retain short pulses after generation is to avoid any added dispersion. In our experiments, we find that it is surprisingly easy to obtain similar RDW emission in the cases of variable and constant pressure by applying a simple scaling rule, though with significant differences in the energy-dependent spectrum of the generated pulses, as previously observed in gas-filled anti-resonant fibers \cite{mak_tunable_2013}. Through numerical simulations, we illuminate the mechanism behind this behaviour. Finally, we study the energy-dependent duration of the RDW pulse. We find that a decreasing pressure gradient reduces pulse stretching during propagation even before the HCF exit, and that widely tuneable transform-limited pulses with \SI{3}{\fs} duration can be generated by using short waveguides.

\begin{figure}
    \includegraphics[width=3.45in]{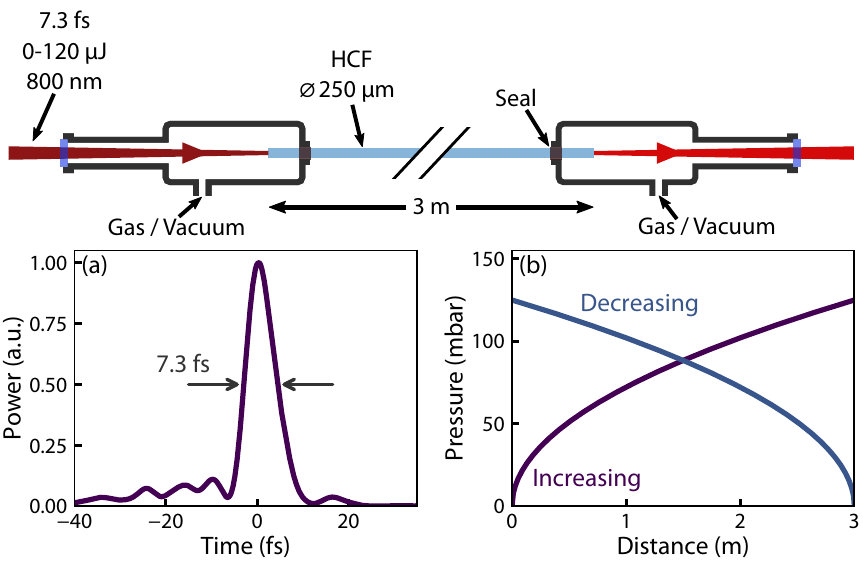}
    \caption{Top: principle of the experiment. \SI{7.3}{\fs} pulses at \SI{800}{\nm} wavelength with energy up to \SI{120}{\micro\joule} are coupled into a \SI{3}{\m} long HCF with \SI{250}{\micro\meter} core diameter, which is sealed into a gas cell at each end.
    Bottom: the driving pulse as measured using SHG-FROG (a) and pressure profiles along the HCF for increasing (purple) and decreasing (blue) gradients with a fill pressure of \SI{125}{\milli\bar} (b).}
    \label{fig:setup}
\end{figure}

The principle of the experimental setup is shown in Fig.~\ref{fig:setup}. It is identical to that detailed in ref.~\cite{travers_high-energy_2019}, except for the added option to create a pressure gradient. In brief: pulses centered at 800~nm with a duration of 30~fs full width at half maximum (FWHM) generated by a titanium-doped sapphire laser amplifier are compressed to 7.3~fs FWHM duration by spectral broadening in a helium-filled HCF (which is stretched to eliminate bend loss \cite{nagy_flexible_2008}) and subsequent phase compensation by reflection off of dispersive mirrors and transmission through silica wedges. The profile of the compressed pulse, measured via second harmonic generation frequency-resolved optical gating (SHG-FROG), is shown in Fig.~\ref{fig:setup}(a). A combination of a motorized half-wave plate and a Brewster-angle silicon plate acts as a broadband variable attenuator. The compressed pulses are then coupled into a second stretched HCF with \SI{250}{\micro\meter} core diameter and a length of 3~m. This HCF is sealed into a gas cell at each end, which can be independently evacuated or filled with gas. At the pressures used in our experiments, a single roughing pump is sufficient to keep the pressure in the evacuated cell below 1~mbar. The output from this second stage is analysed with a calibrated system consisting of an integrating sphere and a fiber-coupled spectrometer.

When one end of the HCF is evacuated, the pressure distribution is given by \cite{suda_generation_2005, mak_tunable_2013}
\begin{equation}
    p_\mathrm{i}(z) = p_\mathrm{max} \sqrt{\frac{z}{L}}\,,\qquad p_\mathrm{d}(z) = p_\mathrm{max} \sqrt{1-\frac{z}{L}}\,,
    \label{eq:pressure_vac}
\end{equation}
where $p_\mathrm{i}(z)$ and $p_\mathrm{d}(z)$ denote an increasing and decreasing pressure gradient, respectively, $z$ is the position along an HCF with total length $L$, and $p_\mathrm{max}$ is the pressure on the high-pressure end [see Fig.~\ref{fig:setup}(b)]. In conventional HCF-based post-compression systems, the dispersion of the waveguide is often very weak, with dispersion lengths many times longer than the HCF \cite{travers_high-energy_2019}. With the nonlinearity dominating, similar spectral broadening can be obtained with constant and variable pressure by matching the integrated nonlinear phase shift (also known as the B-integral). For the increasing gradient in eq.~\ref{eq:pressure_vac}, this is given by \cite{suda_generation_2005}
\begin{equation}
    \phi_\mathrm{nl} = P_0 \int_0^L \! \gamma(z)\, \dd z = P_0\, \gamma_\mathrm{max} \int_0^L \!\!\sqrt{\frac{z}{L}}\, \dd z = \frac{2}{3}P_0 L\, \gamma_\mathrm{max}\,,
    \label{eq:phinl}
\end{equation}
where $P_0$ is the peak power of the initial pulse, $\gamma(z)$ is the nonlinear coefficient---proportional to the nonlinear refractive index $n_2(z)$ and hence to the pressure $p(z)$---and $\gamma_\mathrm{max}$ is its value at the high-pressure end. The factor of $2/3$ is obtained for both decreasing and increasing gradients. In a conventional post-compression system, the fill pressure therefore simply needs to be increased by a factor of $3/2$ as compared to the constant-pressure case, where $\phi_\mathrm{nl} = P_0 L\, \gamma_\mathrm{max}$.

\begin{figure}
    \includegraphics[width=3.46in]{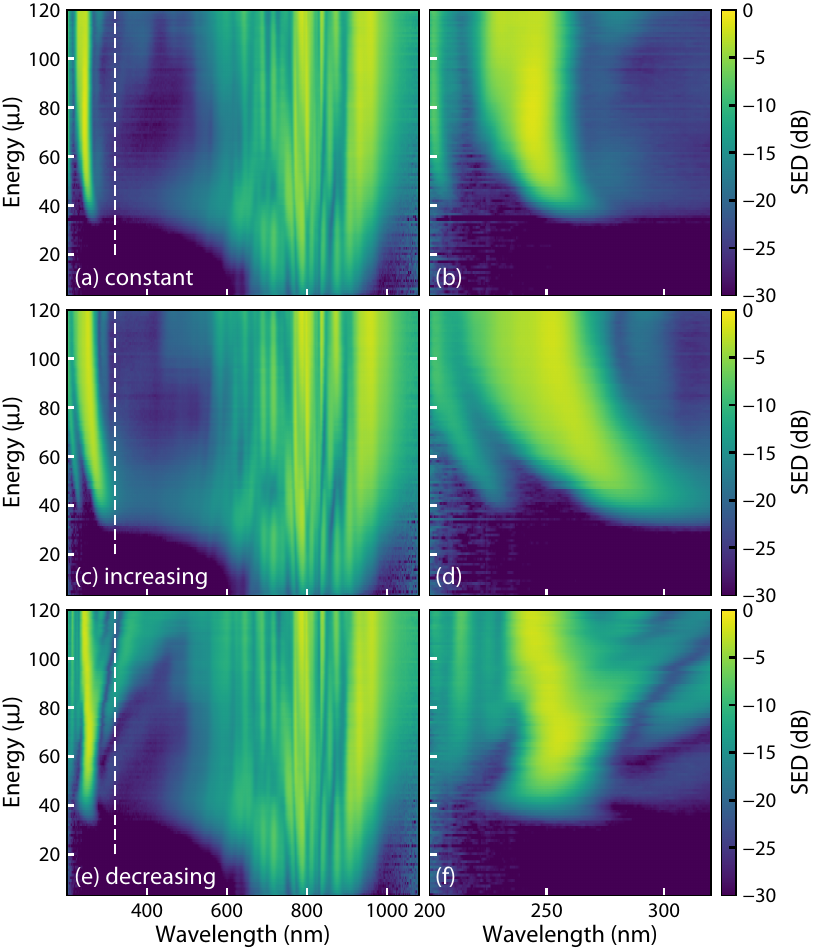}
    \caption{Experimental output spectra (logarithmic color scale) from the second HCF as the driving pulse energy is changed for (a), (b) constant argon pressure of 83~mbar and (c), (d) increasing and (e), (f) decreasing gradients from vacuum to 125~mbar. The right column shows a magnified view of the region between 200~nm and 320~nm [white dashed lines in (a), (c), and (e)]. SED: spectral energy density.}
    \label{fig:data}
\end{figure}

Figure~\ref{fig:data} shows output spectra from the HCF as the input energy is changed for a constant argon pressure of 83~mbar, an increasing gradient from vacuum to 125~mbar, and a decreasing gradient from 125~mbar to vacuum. The constant-pressure data shows typical behaviour observed in many previous studies of soliton self-compression and resonant dispersive wave emission in gas-filled hollow waveguides \cite{joly_bright_2011,brahms_high-energy_2019-1,kottig_generation_2017,mak_tunable_2013,travers_high-energy_2019,brahms_high-energy_2019}: as the energy is increased, the spectrum broadens dramatically, until at some point a dispersive wave is generated at a much shorter wavelength than the driving pulse. In this case, the RDW first appears at an energy of around \SI{40}{\micro\joule} and at 250~nm wavelength.

The pressure for the gradients is chosen as a factor of $3/2$ higher than the constant pressure, as it would be for a post-compression system. In contrast to simple spectral broadening, where the pulse shape does not change significantly during propagation, soliton self-compression is a strongly dynamical process. How the pulse evolves depends critically on the strength of both the nonlinearity and the dispersion, and the simple scaling rule accounts for neither the change in dispersion nor the evolution of the pulse during propagation. Despite this, our experimental data shows broadly similar behaviour for both decreasing and increasing gradients. As shown in Fig.~\ref{fig:data}(c) and (e), the RDW is first generated at nearly the same driving pulse energy and its central wavelength is very close to that in Fig.~\ref{fig:data}(a). This close correspondence is unexpected, and we anticipate that it will be useful in designing vacuum-coupled RDW emission systems, since it allows the use of the simple relations that determine the required HCF and pulse parameters for constant pressure \cite{travers_high-energy_2019,brahms_high-energy_2019}.

Significant differences between the three cases do appear in the energy-dependent RDW spectrum. For constant pressure [Fig.~\ref{fig:data}(b)], the RDW spectrum initially shifts to shorter wavelengths as the energy is increased, but this blue-shift saturates at energies above $\sim\!\SI{60}{\micro\joule}$. In an increasing gradient [Fig.~\ref{fig:data}(d)], the energy-dependent blue-shift is much stronger, covering nearly \SI{50}{\nm} between \SI{40}{\micro\joule} and \SI{80}{\micro\joule} and only saturating around $\sim\!\SI{100}{\micro\joule}$. This could be a useful tool in applications where fast spectral tuneability over a relatively small range is required, since the driving energy can be changed much more quickly than the gas pressure. Finally, for a decreasing gradient [Fig.~\ref{fig:data}(f)], the RDW spectrum initially shifts to longer wavelengths before moving back at high energies. The overall shift is much reduced and comparable to the case of constant pressure. Similar behaviour has been observed in gas-filled anti-resonant fiber when using identical gas pressure for the three cases, leading to different RDW emission wavelengths \cite{mak_tunable_2013}. Here, the similarities in the overall dynamics allow us to more directly compare the evolution of the dispersive wave.

\begin{figure}[t]
    \includegraphics[width=3.46in]{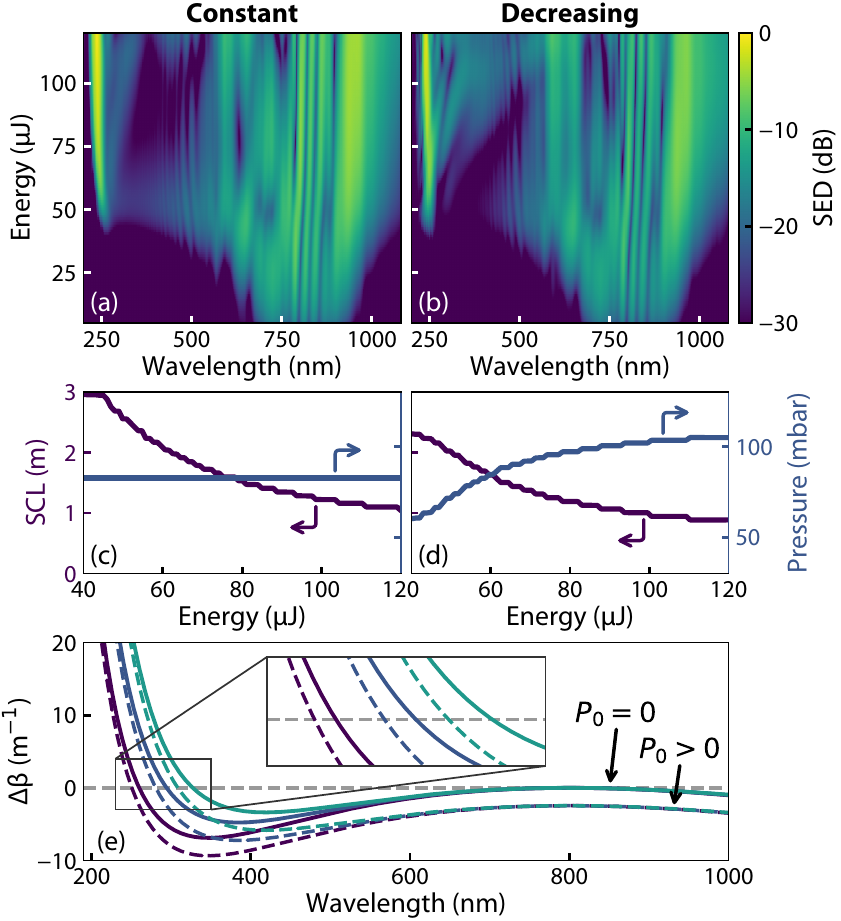}
    \caption{Top row: simulated output spectra, on a logarithmic color scale, for the case of constant pressure (a) and decreasing gradient (b). Middle row: self-compression length (SCL, left axis) and pressure at the SCL (right axis) for the case of constant pressure (c) and decreasing gradient (d). The discrete steps in this data reflect the stepsize in the numerical simulation. Bottom row: dephasing rate between the dispersive wave and soliton as calculated using eqs.~\ref{eq:bsol} and \ref{eq:blin} for an argon-filled \SI{250}{\micro\meter} diameter HCF. The colors correspond to (from dark to light) 80, 100, and 120~mbar pressure. Solid and dashed lines correspond to $P_0 = 0$ and $P_0 = \SI{12.5}{\giga\watt}$, respectively.}
    \label{fig:simulations}
\end{figure}

To examine this behaviour more closely, we simulate the RDW emission process using the model detailed in ref.~\cite{travers_high-energy_2019}. There are no free parameters in this model---we use the measured input pulse shown in Fig.~\ref{fig:setup}(a) and the coupled pulse energy as determined by the transmission of the evacuated HCF. The pressure gradient is described by eq.~\ref{eq:pressure_vac}. Because of its advantages in delivering pulses to an experiment, we focus on comparing a decreasing gradient to the constant-pressure case. The simulated output spectra, shown in Fig.~\ref{fig:simulations}(a) and (b), faithfully reproduce the experimental data in Fig.~\ref{fig:data}(a-b) and (e-f).

At higher driving energy, the stronger nonlinear interaction leads to faster self-compression. As shown in the left axis in Figs.~\ref{fig:simulations}(c) and (d), the self-compression length (SCL), defined here as the distance at which the self-compressing pulse reaches its maximum peak power, becomes much shorter. (Note that in the case of a decreasing gradient, the pulse always reaches its maximum peak power before the end of the HCF, since the near-zero nonlinearity near the exit prevents further self-compression.) The RDW is emitted very close to the self-compression point. In a gradient, this means that it is generated at different pressures depending on the energy [right axis in Fig.~\ref{fig:simulations}(d)]. The different pulse energy and gas pressure then affect the RDW phase-matching.

The wavelength of RDW emission is determined by the phase mismatch between the strongly nonlinear self-compressing soliton and a weaker linear (dispersive) wave. Their wavevectors can be approximated as
\begin{align}
    \beta_\mathrm{s}(\omega) &= \beta_0 + \beta_1 \Delta\omega + \frac{\gamma P_0}{2} \label{eq:bsol}\\
    \beta_\mathrm{l}(\omega) &= \beta_0 + \beta_1 \Delta\omega + \frac{\beta_2}{2} \Delta\omega^2 + \frac{\beta_3}{6} \Delta\omega^3 + \cdots\,,
    \label{eq:blin}
\end{align}
where $\beta_\mathrm{s}$ and $\beta_\mathrm{l}$ are the soliton and linear wavevectors, respectively, and $\Delta\omega = \omega-\omega_0$ is the frequency detuning from the central frequency of the soliton $\omega_0$. The phase mismatch $\Delta\beta = \beta_\mathrm{l} - \beta_\mathrm{s}$ for our experimental parameters is shown in Fig.~\ref{fig:simulations}(e). RDW emission occurs where the soliton and dispersive wave are phase-matched, i.e. $\Delta\beta = 0$. For 80~mbar of argon pressure (purple lines), the phase mismatch is zero at around 260~nm, close to the experimentally measured emission wavelength of 250~nm. At higher pressures (blue and green lines), the amount of anomalous dispersion is decreased and higher-order dispersion increased, leading to phase-matching at longer wavelengths---this is the origin of the wide tuneability observed in constant-pressure experiments \cite{travers_high-energy_2019,brahms_high-energy_2019}. Higher peak power $P_0$ has the opposite effect [dashed lines in Fig.~\ref{fig:simulations}(e)], shifting the dephasing curve down by $\gamma P_0/2$ and leading to shorter phase-matching wavelengths. 

The interaction between the changing self-compression length and the phase-matching explains the behaviour observed in our experiments. With constant pressure, the linear dispersion is also constant, and only the nonlinear shift acts on the spectrum. The resulting blue-shift is commonly observed in studies of RDW emission \cite{joly_bright_2011}. For an increasing gradient, faster self-compression means RDW emission at lower pressures. The nonlinear and linear contributions to the phase-matching thus add up, with both effects pushing the dispersive wave to shorter wavelengths. The resulting strong nonlinear dependence leads to the fast tuning observed in the experiment. In a decreasing gradient, the two effects compete instead, since RDW emission occurs at higher pressure for higher driving energy. Remarkably, these influences nearly cancel each other, so that the RDW spectrum moves by less than \SI{10}{\nm} even as the pulse energy is increased by a factor of 3. These phase-matching considerations are based on propagation purely in the fundamental mode of the HCF. Therefore the RDW is generated with a near-ideal spatial profile, as previously shown experimentally \cite{travers_high-energy_2019}.

\begin{figure}
    \includegraphics[width=3.45in]{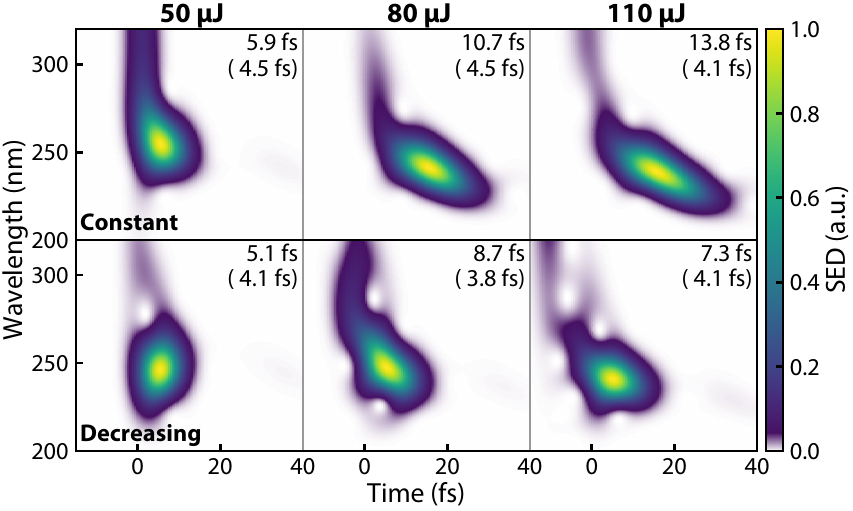}
    \caption{Time-frequency representation of the resonant dispersive wave pulse at the end of the HCF for constant pressure (top) and a decreasing gradient (bottom), extracted from the simulations shown in Fig.~\ref{fig:simulations} for the driving pulse energies indicated at the top of each column. The spectrograms were calculated using a Gaussian gate with a FWHM of 8~fs, and each is normalized to its peak. The labels give the FWHM duration as well as the Fourier-transform limit (in parentheses) of the pulse between 210~nm and 280~nm.}
    \label{fig:spectrograms}
\end{figure}

With our simulations, we can directly compare the output pulse duration achieved with a decreasing gradient and constant pressure. Figure~\ref{fig:spectrograms} shows spectrograms of the RDW pulse at the HCF exit for different driving energies in the two simulations shown in Fig.~\ref{fig:simulations}. In both cases, the shortest pulse is obtained at low energy, with RDW emission occurring close to the HCF exit. For constant pressure, increasing the energy strongly chirps the pulse, stretching it to over three times its transform limit. In contrast, in a decreasing gradient the pulse remains relatively short and chirp-free even at \SI{110}{\micro\joule} of driving pulse energy, where the RDW is emitted after only \SI{1}{\meter} of propagation. Note that the confinement loss at \SI{250}{\nm} in this large-core capillary is so small as to be negligible even though the RDW propagates another \SI{2}{\meter} after being generated. 

\begin{figure}[t]
    \includegraphics[width=3.45in]{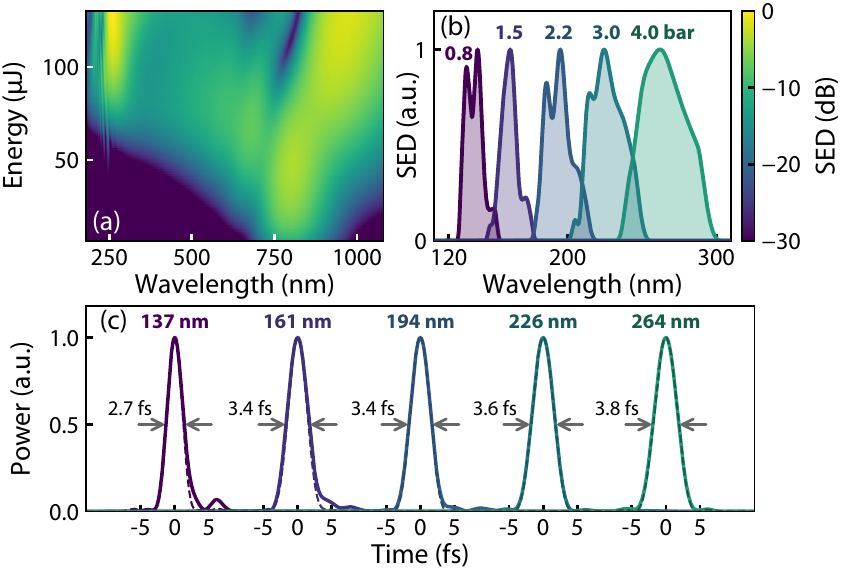}
    \caption{Top: spectral evolution with increasing energy for \SI{4}{\bar} helium pressure (a) and output spectra at the pressures indicated by the labels in units of bar (b) in a \SI{1}{\meter} long, \SI{250}{\micro\meter} core diameter HCF filled with a decreasing gradient and pumped with \SI{7.5}{\fs} pulses of varying energy. Bottom: output pulse duration for the spectra shown in (b) with the central wavelength and FWHM pulse duration indicated. The dashed lines show the transform-limited pulses.}
    \label{fig:short_pulses}
\end{figure}

While the decreasing gradient leads to a dramatic improvement, the duration of the pulses in Fig.~\ref{fig:spectrograms} is significantly longer than previously measured in anti-resonant fibers \cite{brahms_direct_2019}. Our simulations suggest that very broadband and nearly transform-limited pulses at a variety of wavelengths can be obtained by using a shorter HCF and hence further reducing the dispersion experienced by the RDW after generation. Figure~\ref{fig:short_pulses}(a) shows output spectra as the energy is varied from a \SI{1}{\meter} HCF with \SI{250}{\micro\meter} core diameter filled with a decreasing pressure gradient. Here we use a Gaussian input pulse with \SI{7.5}{\fs} duration. The spectral evolution is very similar to that observed in the \SI{3}{\meter} HCF, though more energy is required to achieve RDW emission due to the shorter propagation length. Changing the gas pressure at the HCF entrance tunes the RDW wavelength \cite{mak_tunable_2013,travers_high-energy_2019,brahms_direct_2019}. Figure~\ref{fig:short_pulses}(b) shows output spectra obtained with helium pressures between 0.8 and \SI{4}{\bar}. (The use of helium is necessary to avoid ionisation effects which would occur in argon due to the higher driving energy, especially at low pressure.) The corresponding output pulses, along with their transform-limited shapes, are shown in Fig.~\ref{fig:short_pulses}(c). The driving energy for each pressure is chosen to maximize the peak power of the RDW pulse at the exit, and each spectrum is band-pass filtered with \SI{15}{\percent} relative bandwidth to emulate the effect of filtering optics. At all wavelengths, the output RDW pulse is around \SI{3}{\fs} in duration; above \SI{194}{\nm}, the simulated and transform-limited pulse shapes are virtually indistinguishable.

In summary, we have shown that broadly similar resonant dispersive wave emission can be obtained in HCF filled with either constant gas pressure or gradients by employing a very simple scaling rule of a factor $3/2$ increase in gas pressure. We have clearly identified the interplay between a change in self-compression length and the linear and nonlinear phase-matching contributions as the mechanism behind the different spectral shifts of the dispersive wave that are observed for three pressure profiles. We have furthermore found that a decreasing pressure gradient is not only useful in eliminating the dispersion caused by transmission through windows, but also significantly reduces the pulse stretching within the waveguide itself. Finally, extremely short, transform-limited pulses can be generated by moving to shorter waveguides. We anticipate that our results will be very useful in exploiting RDW-based tuneable few-femtosecond light sources for cutting-edge time-resolved spectroscopy experiments as well as other applications.

\bibliography{bibliography}

\begin{thebibliography}{10}
\newcommand{\enquote}[1]{``#1''}

\bibitem{kotsina_ultrafast_2019}
N.~Kotsina, F.~Belli, S.-F. Gao, Y.-Y. Wang, P.~Wang, J.~C. Travers, and
  D.~Townsend, {\protect\JournalTitle{J. Phys. Chem. Lett.}} \textbf{10}, 715
  (2019).

\bibitem{maiuri_ultrafast_2020}
M.~Maiuri, M.~Garavelli, and G.~Cerullo, {\protect\JournalTitle{J. Am. Chem.
  Soc.}} \textbf{142}, 3 (2020).

\bibitem{ermolov_supercontinuum_2015}
A.~Ermolov, K.~F. Mak, M.~H. Frosz, J.~C. Travers, and P.~S.~J. Russell,
  {\protect\JournalTitle{Phys. Rev. A}} \textbf{92}, 033821 (2015).

\bibitem{travers_high-energy_2019}
J.~C. Travers, T.~F. Grigorova, C.~Brahms, and F.~Belli,
  {\protect\JournalTitle{Nat. Photonics}} \textbf{13}, 547 (2019).

\bibitem{brahms_high-energy_2019-1}
C.~Brahms, F.~Belli, and J.~C. Travers, {\protect\JournalTitle{ArXiv191010458
  Phys.}}  (2019).

\bibitem{joly_bright_2011}
N.~Y. Joly, J.~Nold, W.~Chang, P.~H{\"o}lzer, A.~Nazarkin, G.~K.~L. Wong,
  F.~Biancalana, and P.~S.~J. Russell, {\protect\JournalTitle{Phys. Rev.
  Lett.}} \textbf{106}, 203901 (2011).

\bibitem{brahms_high-energy_2019}
C.~Brahms, T.~Grigorova, F.~Belli, and J.~C. Travers,
  {\protect\JournalTitle{Opt. Lett.}} \textbf{44}, 2990 (2019).

\bibitem{mak_tunable_2013}
K.~F. Mak, J.~C. Travers, P.~H{\"o}lzer, N.~Y. Joly, and P.~S.~J. Russell,
  {\protect\JournalTitle{Opt. Express}} \textbf{21}, 10942 (2013).

\bibitem{nisoli_generation_1996}
M.~Nisoli, S.~De~Silvestri, and O.~Svelto, {\protect\JournalTitle{Appl Phys
  Lett}} \textbf{68}, 2793 (1996).

\bibitem{suda_generation_2005}
A.~Suda, M.~Hatayama, K.~Nagasaka, and K.~Midorikawa,
  {\protect\JournalTitle{Appl Phys Lett}} \textbf{86}, 111116 (2005).

\bibitem{brahms_direct_2019}
C.~Brahms, D.~R. Austin, F.~Tani, A.~S. Johnson, D.~Garratt, J.~C. Travers,
  J.~W.~G. Tisch, P.~S. Russell, and J.~P. Marangos,
  {\protect\JournalTitle{Opt. Lett.}} \textbf{44}, 731 (2019).

\bibitem{nagy_flexible_2008}
T.~Nagy, M.~Forster, and P.~Simon, {\protect\JournalTitle{Appl. Opt.}}
  \textbf{47}, 3264 (2008).

\bibitem{kottig_generation_2017}
F.~K{\"o}ttig, F.~Tani, C.~M. Biersach, J.~C. Travers, and P.~S. Russell,
  {\protect\JournalTitle{Optica}} \textbf{4}, 1272 (2017).

\end{thebibliography}
\bibliographyfullrefs{bibliography}
\end{document}